\documentclass[prb,twocolumn,showpacs,float]{revtex4-1}
\usepackage{graphicx}
\usepackage{amssymb}
\usepackage{amsmath}
\usepackage{xspace}
\usepackage{url}

\newcommand{\phdag}{{\phantom{\dag}}}
\newcommand{\abs}[1]{\left\vert#1\right\vert}

\begin{document} 
\title{Mixed-valence correlations in charge-transferring atom-surface collisions}

\author{M. Pamperin, F. X. Bronold, and H. Fehske}
\affiliation{Institut f{\"ur} Physik,
             Ernst-Moritz-Arndt-Universit{\"a}t Greifswald,
             17489 Greifswald,
             Germany}
\date{\today}
\begin{abstract}
Motivated by experimental evidence for a mixed-valence state to occur 
in the neutralization of strontium ions on gold surfaces we analyze 
this type of charge-transferring atom-surface collision from a many-body 
theoretical point of view using quantum-kinetic equations together with 
a pseudo-particle representation for the electronic configurations of 
the atomic projectile. Particular attention is paid to the temperature 
dependence of the neutralization probability which--experimentally--seems 
to signal mixed-valence-type correlations affecting the charge-transfer
between the gold surface and the strontium projectile. We also investigate
the neutralization of magnesium ions on a gold surface which shows no 
evidence for a mixed-valence state. Whereas for magnesium excellent 
agreement between theory and experiment could be obtained, for strontium 
we could not reproduce the experimental data. Our results indicate 
mixed-valence correlations to be in principle present, but for the model 
mimicking most closely the experimental situation they are 
not strong enough to affect the neutralization process quantitatively. 
\end{abstract}
\maketitle

\section{Introduction}

Charge-exchange between an atomic projectile and a surface plays a central role
in surface science.~\cite{Winter07,Rabalais94,LG90,BN89,Modinos87,YM86} Many 
surface diagnostics, for instance, secondary ion mass spectrometry~\cite{CH91} 
or meta-stable atom de-excitation spectroscopy~\cite{HMO97} utilize 
surface-based charge-transfer processes. The same holds for plasma science. 
Surface-based production of negative hydrogen ions, for instance, is currently 
considered as a pre-stage process in neutral gas heating of fusion 
plasmas.~\cite{KFF07} The operation modii of low-temperature gas discharges~\cite{LL05},
which are main work horses in many surface modification and semiconductor industries, 
depend on secondary electron emission from the plasma walls
and thus also on surface-based charge-transfer processes.

Besides of their great technological importance, charge-transferring atom-surface collisions
are however also of fundamental interest. This type of collision couples a local quantum
system with a finite number of discrete states--the projectile--to a large reservoir with 
a continuum of states--the target. Irrespective of the coupling between the two, either due 
to tunneling or due to Auger-type Coulomb interaction, charge-transferring atom-surface collisions 
are thus perfect realizations of time-dependent quantum impurity systems.~\cite{SNL96,MM98} 
By a judicious choice of the projectile-target combination as well as the collision parameters
Kondo-type features~\cite{Hewson93} are thus expected as in any other quantum impurity 
system.~\cite{GD92,WM94,AL03,THS09}

Indeed a recent experiment by He and Yarmoff~\cite{HY10,HY11} provides strong evidence 
for electron correlations affecting the neutralization of positively charged strontium ions 
on gold surfaces. The fingerprint of correlations could be the experimentally found negative 
temperature dependence of the neutralization probability. It may arise~\cite{SNL96,MM98}
from thermally excited conduction band holes occupying the strongly renormalized $5s^1$ 
configuration of the projectile which effectively stabilizes the impinging  
ion and reduces thereby the neutralization probability. 
The purpose of the present work is to analyze the He-Yarmoff experiment~\cite{HY10,HY11} 
from a genuine many-body theoretical point of view, following the seminal work of Nordlander 
and coworkers~\cite{LN91,NSL93,SLN94a,SLN94b,SNL96} as well as Merino and Marston~\cite{MM98} 
and to provide theoretical support for the interpretation of the experiment in terms of a 
mixed-valence scenario.

We couch--as usual--the theoretical description of the charge-transferring atom-surface collision 
in a time-dependent Anderson impurity model.~\cite{LG90,BN89,Modinos87,YM86,KO87,NKO88,RFG09,BFG07,GFM05,OM96,MAB93}
The parameters of the model are critical. To be as realistic as possible 
without performing an expensive
ab-initio analysis of the ion-surface interaction 
we employ for the calculation of the model 
parameters  Gadzuk's semi-empirical approach~\cite{Gadzuk67a,Gadzuk67b} based on 
image charges and Hartree-Fock wave functions for the projectile states.~\cite{CR74} The 
time-dependent Anderson model, written in terms of pseudo-operators~\cite{Coleman84,KR86}
for the projectile states, is then subjected to a full quantum-kinetic analysis using 
contour-ordered Green functions~\cite{KB62,Keldysh65} and a non-crossing approximation 
for the hybridization self-energies as originally proposed by Nordlander and 
coworkers.~\cite{LN91,NSL93,SLN94a,SLN94b,SNL96}

We apply the formalism to analyze, respectively, the neutralization of a strontium and a magnesium 
ion on a gold surface. For the Mg:Au system, which shows no evidence for mixed-valence correlations 
affecting the charge-transfer between the surface and the projectile, we find excellent agreement
between theory and experiment. For the Sr:Au system, in contrast, we could reproduce only the 
correct order of magnitude of the  neutralization probability. Its temperature dependence  
could not be reproduced. Our modeling shows however that a mixed-valence scenario could in 
principle be at work. For the material parameters best suited for the description
of the Sr:Au system they are however not strong enough to affect the neutralization probability 
also quantitatively.

The outline of our presentation is as follows. In the next section we describe the time-dependent 
Anderson model explaining in particular how we obtained the parameters characterizing it.
Section \ref{QuantumKinetics} concerns the quantum kinetics and presents the set of coupled 
two-time integro-differential equations which have to be solved for determining the probabilities 
with which the various charge states of the projectile occur. They form the basis for the 
analysis of the temperature dependence of the neutralization probability. Numerical results 
for a strontium as well as a magnesium ion hitting a gold surface are presented, discussed, 
and compared to experimental data in Sect.~\ref{Results}. Concluding remarks are given 
in Sect.~\ref{Conclusions}.

\section{Model}
\label{Model}

When an atomic projectile approaches a surface its energy levels shift and broaden due to 
direct and exchange Coulomb interactions with the surface. Since the target and the 
projectile are composite objects the calculation of these shifts and broadenings from first 
principles is a complicated problem.~\cite{NT90} We follow therefore Gadzuk's semi-empirical 
approach.~\cite{Gadzuk67a,Gadzuk67b} From our previous work on secondary electron emission 
due to de-excitation of meta-stable nitrogen molecules on metal~\cite{MBF11} and  
dielectric~\cite{MBF12a,MBF12b} surfaces we expect the approach to give reasonable estimates
for the level widths as well as the level positions for distances from the surface larger 
than a few Bohr radii. In addition, the approach has a clear physical picture behind it and 
is thus intuitively very appealing.

\begin{figure}[t]
\includegraphics[width=0.8\linewidth]{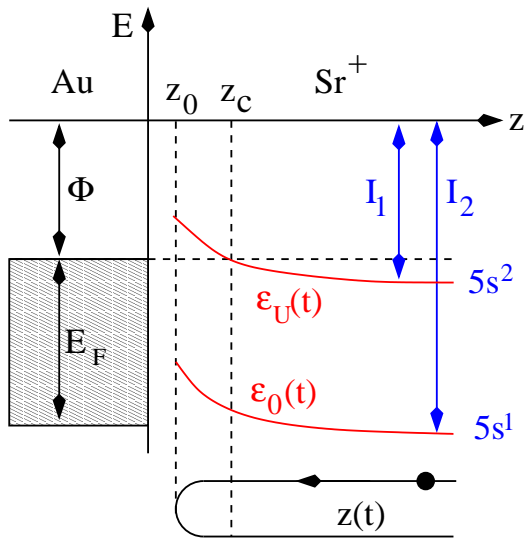}
\caption{Illustration of the time-dependent quantum impurity model used for
the description of the charge-transferring scattering of a ${\rm Sr}^+$ ion
on a gold surface. The two ionization energies, $\varepsilon_U(t)$ and 
$\varepsilon_0(t)$, standing for the projectiles' $5s^2$ and $5s^1$ configuration, 
respectively, shift due to the image interaction with the surface. Far away 
from the surface the two energies merge, respectively, with the first 
$(I_1)$ and the second ($I_2$) ionization energy of a strontium atom.
The image interaction also leads to a hybridization
of the Sr states with the conduction band states of the
surface which is characterized by a step potential at $z=0$ whose depth
is the sum of the work function $\Phi>0$ and the Fermi energy ${\rm E_F}>0$.
For simplicity the broadening is not shown. Indicated however is the trajectory 
$z(t)$ of the ion. Important points along the trajectory are $z_0$, the turning
point, and $z_c$, the point where the first ionization level crosses the
Fermi energy.}
\label{ANM}
\end{figure}

The essence of the model is illustrated in Fig.~\ref{ANM}. It shows for the
particular case of a strontium ion hitting a gold surface the energy 
levels of the projectile closest to the Fermi energy of the target. Quite generally, 
for alkaline-earth (AE) ions the first and the second ionization levels are 
most important. Identifying the positive ion (${\rm AE}^+$) with a singly occupied 
impurity and the neutral atom (${\rm AE}^0$) with a doubly occupied impurity,
the projectile can be modelled as a non-degenerate, asymmetric Anderson model 
with on-site energies
\begin{align}
\varepsilon_U(z)&=-I_1+\frac{e^2}{4|z-z_i|}~,
\\
\varepsilon_0(z)&=-I_2+\frac{3e^2}{4|z-z_i|}~,
\end{align}
where $I_1>0$ and $I_2>0$ are, respectively, the first and second ionization energy 
far away from the surface while $z_i$ is the distance of the metal's image plane from 
its crystallographic ending at $z=0$. The on-site Coulomb repulsion $U(z)$ would 
be the difference of the two energies. Table~\ref{Parameters} summarizes the material
parameters required for the modeling of the neutralization of strontium and magnesium 
ions on a gold surface.
\begin{table}[b]
\begin{center}
  \begin{tabular}{c|c|c|c|c|c|c|c|c}
           & $I_1$[eV] & $Z_1$ & $I_2$[eV] & $Z_2$ & $\Phi$[eV] & ${\rm E_F}$[eV] & $z_{\rm i}$[a.u.] & $m_e^*/m_e$\\\hline
  {\rm Sr} & 5.7 & 1.65 & 11.0 & 2 & -- & -- & -- & --\\
  {\rm Mg} & 7.65 & 1.65 & 15.04 & 2 & -- & -- & --& -- \\
  {\rm Au} & -- & -- & -- & -- & 5.1--5.2 & 5.53 & 1.0 & 1.1\\
  \end{tabular}
  \caption{Material parameters for magnesium, strontium and gold: $I_1$ and $I_2$ are the 
  first and the second ionization energy, $Z_1$ and $Z_2$ are the effective charges to be 
  used in the calculation of the hybridization matrix element (viz: Eq.~\eqref{VC}), 
  $\Phi$ is the work function, ${\rm E_F}$ the Fermi energy, $z_i$ the position of 
  the image plane in front of the surface, and $m_e^*$ is the effective mass of an electron.}
  \label{Parameters}
\end{center}
\end{table}

The $z-$dependent shifts of the ionization levels can be obtained as the energy 
gain of a virtual process moving the configuration under consideration from the 
actual position $z$ to $z=\infty$, reducing its electron occupancy by one, and then 
moving it back to position $z$, taking into account in both moves--if present--image 
interactions due to the charge state of the final and initial configurations with 
the metal.~\cite{NMB83} For the upper level, $\varepsilon_U$, that is, the $5s^2$ 
configuration the cycle is ${\rm AE} \rightarrow {\rm AE}^+ + e^- \rightarrow {\rm AE}^+$, 
whereas for the lower level, $\varepsilon_0$, that is, the $5s^1$ configuration the 
cycle is 
${\rm AE}^+ \rightarrow {\rm AE}^{2+} + e^- \rightarrow {\rm AE}^{2+}$.

To set up the Hamiltonian we also need the wave functions for the projectile states. 
For the upper level we use the (ns) Hartree-Fock wave function of an ${\rm AE}$ atom
while for the lower level we use the (ns) Hartree-Fock wave function of an ${\rm AE}^+$ ion.
According to Clementi and Roetti~\cite{CR74} both can be written in the form 
\begin{align}
\psi_{\rm HF}(\vec{r}\,)&=Y_{00}(\theta, \phi) \sum_{j=1}^N c_j N_j |\vec{r}\,|^{n_j-1}
e^{-C_j |\vec{r}\,|}~
\end{align}
with $c_j, n_j, C_j$, and $N_j$ tabulated parameters and $Y_{00}(\theta, \phi)$ the 
spherical harmonics with $m=l=0$.
\begin{figure}[t]
\includegraphics[width=0.9\linewidth]{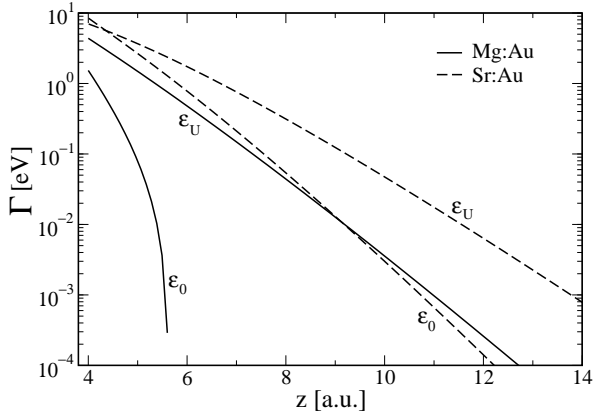}
\caption{Level widths as obtained from Eq.~(\ref{Gamma}) for the
Mg:Au (solid lines) and the Sr:Au (dashed lines) system.}
\label{LevelWidths}
\end{figure}

For simplicity we assume the projectile to approach the surface from $z=\infty$ on a perpendicular
trajectory, 
\begin{align}
z(t)=z_0+v|t|~,
\end{align}
with the turning point $z_0$ reached at time $t=0$ and $v$ the velocity of the
projectile. The lateral motion of the projectile is thus ignored. To be 
consistent with this simple trajectory we also neglect the lateral variation 
of the potential characterizing the metal surface. 
The electrons of the metal are thus simply described in terms of a potential step
at $z=0$ with depth $-|V_0|=\Phi+{\rm E_F}$, where $\Phi>0$ is the work function of the metal
and ${\rm E_F}>0$ is its Fermi energy measured from the bottom of the conduction band
(see Table~\ref{Parameters}), 
leading to 
\begin{align}
        \varepsilon_{\vec{k}} & = \frac{\hbar^2}{2 m^*_e} \left( k_x^2 + k_y^2 + k_z^2 \right)-\abs{V_0}~,
        \\
        \psi_{\vec{k}}(\vec{r}\,) & = \frac{1}{L \sqrt{L}} \, e^{i \left( k_x x + k_y y \right)} 
        \Bigl\{ T_{k_z} e^{-\kappa_{k_z} z} \Theta(z) \nonumber \\
        & + \left[ e^{i k_z z} + R_{k_z} e^{-i k_z z} \right] \Theta(-z) \Bigr\}~,
\end{align}
for the energies and wave functions of the conduction band electrons; $L$ is the spatial width of the step 
(drops out in the final expressions) and 
\begin{align}
        R_{k_z} &= \frac{i k_z + \kappa_{k_z}}{i k_z - \kappa_{k_z}}~,\\
        T_{k_z} &= \frac{2 i k_z}{i k_z - \kappa_{k_z}}~,
\end{align}
with $\kappa_{k_z}=\sqrt{2 m^*_e(\abs{V_0}-k_z^2)/\hbar^2}$ are the reflection and transmission coefficients 
of the potential step. 

While the projectile is on its trajectory its ionization levels hybridize with the conduction band. 
The matrix element for this process is given by~\cite{Gadzuk67a,Gadzuk67b}
\begin{align}
V_{\vec{k}}(t)=\int_{z>0} \!\!\!\!\! d^3r \, \psi_{\vec{k}}^{*}(\vec{r}\,)\frac{Ze^2}{|\vec{r}-\vec{r}_p(t)|}
\psi_{\rm HF}({\vec{r}-\vec{r}_p(t)})~,
\label{VC}
\end{align}
where the potential between the two wave functions is the residual Coulomb interaction of the valence electron with the 
core of the projectile located at $\vec{r}_p(t)=z(t)\vec{e}_z$. The matrix element can be transformed to a level width 
\begin{align}
\Gamma_{\varepsilon(t)}(t)=2\pi\sum_{\vec{k}}\left|V_{\vec{k}}(t)\right|^2\delta(\varepsilon(t)-\varepsilon_{\vec{k}})
\label{Gamma}
\end{align}
which is an important quantity. The charge $Z$ 
in Eq.\eqref{VC} is the charge of the nucleus screened by all the electrons of the projectile except 
of the valence electron under consideration. For the hybridization of the lower level, the second
ionization level, $Z=2$ while for the hybridization of the upper level, the first ionization level, 
$Z=2-s$, where $s=0.35$ is Slater's shielding constant due to the second electron in the 
$s$-valence shell.~\cite{Slater30}

In Fig.~\ref{LevelWidths} we show the level widths calculated form Eq.~\eqref{Gamma} with 
$\varepsilon(t)$ set, respectively, to $\varepsilon_U(t)$ and $\varepsilon_0(t)$, for magnesium 
and strontium using the parameters of Table~\ref{Parameters}. Most probably we overestimate the 
widths close to the surface. To what extent, however, only precise calculations of the kind 
performed for alkaline ions by Nordlander and Tully can show.~\cite{NT90} 
\begin{figure}[t]
\includegraphics[width=0.7\linewidth]{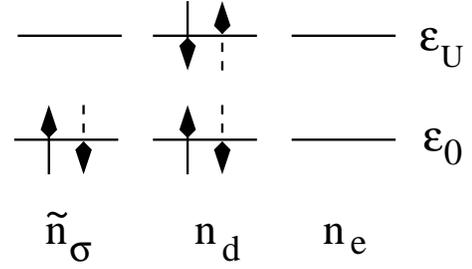}
\caption{Possible configurations of the AE projectile. Solid and
dashed arrows indicate, respectively, spin-reversed states which are
energetically degenerate. The quantities $\tilde{n}_\sigma$,
$n_d$, and $n_e$ are, respectively, the (pseudo) probabilities with which the
${\rm AE}^+$, the ${\rm AE}^0$, and the ${\rm AE}^{2+}$ configuration occur.}
\label{PseudoStates}
\end{figure}

Using Coleman's pseudo-particle representation~\cite{Coleman84,KR86} for the projectile configurations
illustrated in Fig.~\ref{PseudoStates}, the Hamiltonian describing the interaction of an AE
projectile with a metal surface can be written as~\cite{SLN94b} 
\begin{align}
H(t) &= \sum_\sigma \varepsilon_0(t)p_\sigma^\dagger p_\sigma^\phdag 
+ [\varepsilon_0(t)+\varepsilon_U(t)]d^\dagger d^\phdag \nonumber\\
&+\sum_{\vec{k}\sigma}\varepsilon_{\vec{k}} c_{\vec{k}\sigma}^\dagger c_{\vec{k}\sigma}^\phdag
+\sum_{\vec{k}\sigma}\big[V_{\vec{k}}(t)c_{\vec{k}\sigma}^\dagger e^\dagger p_\sigma^\phdag 
 + h.c. \big]\nonumber\\
&+\sum_{\vec{k}\sigma}\big[V_{\vec{k}}(t)c_{\vec{k}\sigma}^\dagger d^\phdag p_{-\sigma}^\dagger + 
h.c. \big]
\label{Hamiltonian}
\end{align}
with $e^\dagger$, $d^\dagger$, and $p_\sigma^\dagger$ denoting, respectively, the creation 
operators for an empty (${\rm AE}^{2+}$), a doubly occupied (${\rm AE}^{0}$), and a singly 
occupied (${\rm AE}^{+}$) projectile. Since the projectile can be only in either one of these 
configurations, the Hamiltonian has to be constrained by~\cite{Coleman84,KR86}
\begin{align}
Q=\sum_\sigma p_\sigma^\dagger p_\sigma^\phdag + d^\dagger d^\phdag 
+ e^\dagger e^\phdag = 1~.
\label{Constraint}
\end{align}

This completes the description of the model. Combined with measured projectile 
velocities the model describes the charge-transfer responsible for the neutralization of 
alkaline-earth ions on noble metal surfaces. 
\begin{figure}[t]
\includegraphics[width=0.8\linewidth]{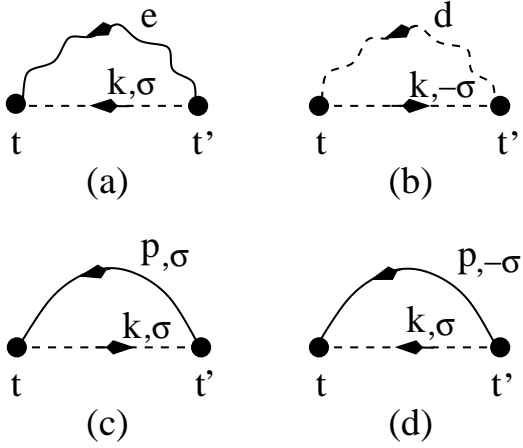}
\caption{Self-energies in the non-crossing approximation. Straight dashed lines
denote bare Green functions for the conduction band electrons. The other
lines indicate renormalized Green functions for the singly occupied
(p), the empty (e), and the doubly occupied (d) AE projectile. Filled
bullets stand for the hybridization matrix element $V_{\vec{k}}(t)$.
Diagrams (a) and (b) give, respectively, the self-energies $\Sigma_{0,\sigma}$
and $\Sigma_{U,\sigma}$ for the Green function $P_\sigma$. The self-energies
$\Pi_{e}$ and $\Pi_{d}$ for the Green functions $E$ and $D$, respectively,
are shown in (c) and (d).}
\label{NCA}
\end{figure}

\section{Quantum kinetics}
\label{QuantumKinetics}

To calculate the neutralization probability for the AE ion hitting the metal 
surface we follow Nordlander and coworkers~\cite{LN91,NSL93,SLN94a,SLN94b,SNL96} and 
set up quantum-kinetic equations for contour-ordered Green functions~\cite{KB62,Keldysh65} 
describing the empty, singly, and doubly occupied projectile. We denote these functions, 
respectively, by $E(t,t^\prime)$, $P_\sigma(t,t^\prime)$, and $D(t,t^\prime)$ and write
their analytic pieces in the form 
\begin{align}
H^{\rm R}(t,t^\prime) &=-i\Theta(t-t^\prime)
                       \exp[-i\!\int_{t^\prime}^t\!\! d\bar{t}\varepsilon(\bar{t})] \bar{H}^{\rm R}(t,t^\prime)~,\\
H^{\gtrless}(t,t^\prime) &=\exp[-i\!\int_{t^\prime}^t \!\!d\bar{t}\varepsilon(\bar{t})] \bar{H}^{\gtrless}(t,t^\prime)~,
\end{align}
where $H(t,t^\prime)$ can be any of the three Green functions and $\varepsilon(t)$ is, depending 
on the Green function, either identically $0$, $\varepsilon_0(t)$, or $\varepsilon_0(t)+\epsilon_U(t)$.

Using this notation and calculating the self-energies $\Pi_{e}$, $\Sigma_{0,\sigma}$, $\Sigma_{U,\sigma}$, 
and $\Pi_{d}$ in the non-crossing approximation diagrammatically shown in Fig.~\ref{NCA} leads after 
application of the Langreth-Wilkins rules~\cite{LW72} and the projection to the 
$Q=1$ subspace~\cite{LN91,AL03} to~\cite{SLN94a,SLN94b}
\begin{align}
\frac{\partial}{\partial t} \bar{E}^{\rm R}(t,t^\prime) &= -\sum_\sigma \int_{t^\prime}^t \!\!d\bar{t}
\bar{K}_{\varepsilon_0}^{<}(\bar{t},t)\bar{P}_\sigma^{\rm R}(t,\bar{t})\bar{E}^{\rm R}(\bar{t},t^\prime)~,
\label{Er}\\
\frac{\partial}{\partial t} \bar{P}_\sigma^{\rm R}(t,t^\prime) &= -\int_{t^\prime}^t \!\!d\bar{t}
\bar{K}_{\varepsilon_0}^{>}(t,\bar{t})\bar{E}^{\rm R}(t,\bar{t})\bar{P}_\sigma^{\rm R}(\bar{t},t^\prime)
\label{Pr}\nonumber\\
&~~~-\int_{t^\prime}^t \!\!d\bar{t}
\bar{K}_{\varepsilon_U}^{<}(\bar{t},t)\bar{D}^{\rm R}(t,\bar{t})\bar{P}_\sigma^{\rm R}(\bar{t},t^\prime)~,
\\
\frac{\partial}{\partial t} \bar{D}^{\rm R}(t,t^\prime) &= -\sum_\sigma \int_{t^\prime}^t \!\!d\bar{t}
\bar{K}_{\varepsilon_U}^{>}(t,\bar{t})\bar{P}_{-\sigma}^{\rm R}(t,\bar{t})\bar{D}^{\rm R}(\bar{t},t^\prime)~,
\label{Dr}
\end{align}
and 
\begin{widetext}
\begin{align}
\frac{\partial}{\partial t} \bar{E}^{<}(t,t^\prime) &=\sum_\sigma\int^{t^\prime}_{-\infty} \!\!d\bar{t}
\bar{K}_{\varepsilon_0}^{>}(\bar{t},t)\bar{P}_\sigma^{<}(t,\bar{t})[\bar{E}^{\rm R}(t^\prime,\bar{t})]^*
-\sum_\sigma\int^{t}_{-\infty} \!\!d\bar{t}
\bar{K}_{\varepsilon_0}^{<}(\bar{t},t)\bar{P}_\sigma^{\rm R}(t,\bar{t})\bar{E}^{<}(\bar{t},t^\prime)~,
\label{Elt}\\
\frac{\partial}{\partial t} \bar{P}_\sigma^{<}(t,t^\prime) &=\int^{t^\prime}_{-\infty} \!\!d\bar{t}
\bar{K}_{\varepsilon_0}^{<}(t,\bar{t})\bar{E}^{<}(t,\bar{t})[\bar{P}_\sigma^{\rm R}(t^\prime,\bar{t})]^*
+\int^{t^\prime}_{-\infty} \!\!d\bar{t}
\bar{K}_{\varepsilon_U}^{>}(\bar{t},t)\bar{D}^{<}(t,\bar{t})[\bar{P}_\sigma^{\rm R}(t^\prime,\bar{t})]^*
\nonumber\\
&-\int^{t}_{-\infty} \!\!d\bar{t}
\bar{K}_{\varepsilon_0}^{>}(t,\bar{t})\bar{E}^{\rm R}(t,\bar{t})\bar{P}_\sigma^{<}(\bar{t},t^\prime)
-\int^{t}_{-\infty} \!\!d\bar{t}
\bar{K}_{\varepsilon_U}^{<}(\bar{t},t)\bar{D}^{\rm R}(t,\bar{t})\bar{P}_\sigma^{<}(\bar{t},t^\prime)~,
\label{Plt}\\
\frac{\partial}{\partial t} \bar{D}^{<}(t,t^\prime) &= \sum_\sigma\int^{t^\prime}_{-\infty} \!\!d\bar{t}
\bar{K}_{\varepsilon_U}^{<}(t,\bar{t})\bar{P}_{-\sigma}^{<}(t,\bar{t})[\bar{D}^{\rm R}(t^\prime,\bar{t})]^*
-\sum_\sigma\int^{t}_{-\infty} \!\!d\bar{t}
\bar{K}_{\varepsilon_U}^{>}(t,\bar{t})\bar{P}_{-\sigma}^{\rm R}(t,\bar{t})\bar{D}^{<}(\bar{t},t^\prime)~
\label{Dlt}
\end{align}
\end{widetext}
with 
\begin{align}
\bar{K}_\varepsilon^{\gtrless} (t,t')=\sqrt{\Gamma_{\varepsilon(t)}(t)\Gamma_{\varepsilon(t^\prime)}(t^\prime)}
\bar{f}_{\varepsilon}^{\,\gtrless} (t,t')
\label{Kfct}
\end{align}
and
\begin{align}
\bar{f}_\varepsilon^{\,\gtrless} (t,t')=\exp[i\!\int_{t^\prime}^t\!\! d\bar{t}\varepsilon(\bar{t})]
                                        f^{\,\gtrless}(t-t^\prime)~,
\end{align}
where $f^<(t)=1-f^>(t)$ is the Fourier transform of the Fermi function $f^<(\varepsilon)$ defined by 
\begin{align}
f^<(t)=\int\frac{d\varepsilon}{2\pi} f^<(\varepsilon)\exp[-i\varepsilon t]~. 
\end{align}

The function $\bar{K}_{\varepsilon}^{\gtrless} (t,t')$, which contains the temperature dependence,
entails an approximate momentum summation. From the diagrams shown in Fig.~\ref{NCA} one initially 
obtains
\begin{align}
K^{\gtrless}(t,t')=\int \!\frac{d\varepsilon}{2\pi}
\sqrt{\Gamma_\varepsilon(t)\Gamma_\varepsilon(t^\prime)}
f^{\gtrless}(\varepsilon)\exp[-i\varepsilon(t-t^\prime)]
\end{align}
with an energy integration extending over the range of the conduction band and 
$\Gamma_\varepsilon(t)$ given by Eq.~\eqref{Gamma} with $\varepsilon(t)$ replaced by the
integration variable $\varepsilon$. To avoid the numerically costly energy integration
Nordlander and coworkers employed two different approximations: In Ref.~\onlinecite{SLN94a} 
they replaced $\Gamma_\varepsilon(t)$ by an average over the energy range of the conduction 
band while in Ref.~\onlinecite{SLN94b} they replaced it by $\Gamma_{\varepsilon(t)}(t)$ with 
$\varepsilon(t)$ set to $\varepsilon_0(t)$ or $\varepsilon_U(t)$ depending on which state 
is considered in the hybridization self-energy. Using the latter leads to 
\begin{align}
K_{\varepsilon}^\gtrless(t,t')\simeq
\sqrt{\Gamma_{\varepsilon(t)}(t)\Gamma_{\varepsilon(t^\prime)}(t^\prime)}
f^\gtrless(t-t^\prime)
\end{align}
and eventually to $\bar{K}_{\varepsilon}^{\gtrless}(t,t')$ as given in Eq.~\eqref{Kfct}.
The subscript $\varepsilon$ indicates now not an integration variable but the functional 
dependence on $\varepsilon(t)$. We employ this form but keep in mind that it is an approximation 
to the non-crossing self-energies.

The instantaneous (pseudo) occurrence probabilities  
for the projectile configurations ${\rm AE}^{2+}$, ${\rm AE}^{+}$, and ${\rm AE}^0$ are 
then given by
\begin{align}
n_e(t) &=\bar{E}^<(t,t)~,\\
\tilde{n}_\sigma(t) &=\bar{P}^<_\sigma(t,t)~,\\
n_d(t) &=\bar{D}^<(t,t)~,
\end{align}
respectively, where we refer to all of them as (pseudo) occurrence probabilities also
strictly speaking $n_d$ and $n_e$ are true ones and only 
$\tilde{n}_\sigma$ is a pseudo occurrence probability in the sense that the 
true probability with which the ${\rm AE}^+$ configuration occurs is 
$n_\sigma=\tilde{n}_\sigma+n_d$.~\cite{SLN94b} Sometimes we will also
refer to $n_e$, $\tilde{n}_\sigma$, and $n_d$ simply as (pseudo) occupancies.
For the AE ion the probability for neutralization at the surface (wall recombination) 
is the probability for double occupancy after the completion of the trajectory, that is,
\begin{align}
\alpha_w=n_d(\infty)~,
\end{align}
subject to the initial conditions $n_d(-\infty)=n_e(-\infty)=0$ and $\tilde{n}_\sigma(-\infty)=\delta_{\sigma,1/2}$.

We solve the two coupled sets of integro-differential equations \eqref{Er}--\eqref{Dr} and \eqref{Elt}--\eqref{Dlt}
on a two-dimensional time grid setting 
\begin{align}
\bar{E}^{\rm R}(t,t)=\bar{P}_\sigma^{\rm R}(t,t)=\bar{D}^{\rm R}(t,t)=1
\label{BCr}
\end{align}
for the retarded Green functions and 
\begin{align}
E^<(-\infty,-\infty)&=n_e(-\infty)=0~,
\label{BCelt}\\
P_\sigma^<(-\infty,-\infty)&=\tilde{n}_\sigma(-\infty)=\delta_{\sigma,1/2}~,
\label{BCplt}\\
D^<(-\infty,-\infty)&=n_d(-\infty)=0
\label{BCdlt}
\end{align}
for the less-than Green functions using basically the same numerical strategy as 
Shao and coworkers.~\cite{SLN94a,SLN94b} 
\begin{figure}[t]
\includegraphics[width=0.7\linewidth]{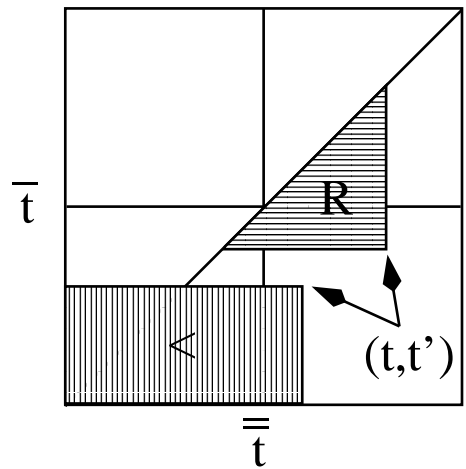}
\caption{Sketch of the domains in the $(\bar{\bar{t}},\bar{t})$ plane 
over which Eqs.(\ref{Er})--(\ref{Dr}) and (\ref{Elt})--(\ref{Dlt}) have 
to be integrated subject to the boundary conditions (\ref{BCr}) and
(\ref{BCelt})--(\ref{BCdlt}), respectively, in order to determine the 
retarded and less-than Green functions at $(t,t^\prime)$. The triangular 
(rectangular) region denotes the domain required for the calculation of
the retarded (less-than) Green functions.}
\label{IntDomain}
\end{figure}

Due to the intertwining of the time integrations the integration domains for the 
retarded Green functions are triangular whereas for the less-than Green function they 
are rectangular as shown in Fig.~\ref{IntDomain}. The size of the time-grid as well as 
the discretization depend on the velocity of the projectile and the maximum distance 
it has from the surface. For the He-Yarmoff experiment the velocities are on the order of 
$0.01$ in atomic units. The maximum distance from which the ion starts its journey can be
taken to be 20 Bohr radii. At this distance the coupling between the surface and the ion is 
vanishingly small. We empirically found the algorithm to converge for a $N\times N$ grid
with $N=1000-3000$. Since the Green functions are complex the computations are time and 
memory consuming. 

\section{Results}
\label{Results}

We now analyze the He-Yarmoff experiment~\cite{HY10,HY11} quantitatively from a many-body 
theoretical point of view. For that purpose we combine the model developed in Sect.~\ref{Model}
with the quantum-kinetics described in Sect.~\ref{QuantumKinetics}. Besides the parameters
given in Table~\ref{Parameters} we also need the velocity of the projectile. In general, 
the velocity will be different on the in- and outgoing branch of the trajectory. The 
outgoing branch, however, determines the final charge state of the projectile. We take 
therefore--for both branches--the normal component of the experimentally measured 
post-collision velocity. If not noted otherwise all quantities are in atomic units, that is, 
energies are measured in Hartrees and lengths in Bohr radii. 
\begin{figure}[t]
\includegraphics[width=0.9\linewidth]{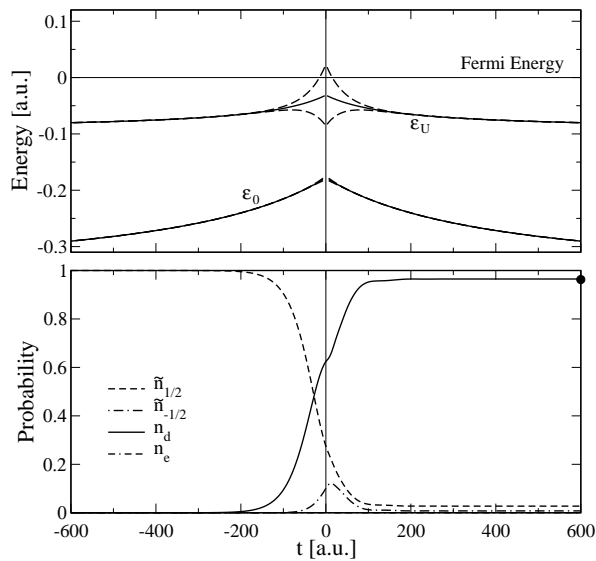}
\caption{Upper panel: Energy level diagram for the Mg:Au system at $T_s=400~$K 
as a function of time. The projectile starts at $t=-600$ and $z=z_{\rm max}=20$ with 
velocity $v=0.024$, reaches at $t=0$ the turning point $z=z_0=5$, and approaches at 
$t=600$ again $z_{\rm max}$. The ionization levels (solid lines) are broadened according 
to $\epsilon_{0,U}\pm \Gamma_{0,U}$ (dashed lines) with $\Gamma_{0,U}$ as shown in 
Fig.~\ref{LevelWidths}. Lower Panel: Instantaneous (pseudo) occurrence probabilities 
along the trajectory for the ${\rm Mg}^+$, the ${\rm Mg^0}$, and the ${\rm Mg}^{2+}$
configurations. Initially, at time $t=-600$, the projectile is in
the ${\rm Mg}^+$ configuration. The neutralization probability
in this particular case is $\alpha_w=n_d(600)=0.965$ (solid bullet).}
\label{OccurrenceMg}
\end{figure}

First, we discuss the Mg:Au system. In Fig.~\ref{OccurrenceMg} we show the time-dependence 
of the broadened ionization levels, $\varepsilon_U$ and $\varepsilon_0$, together with the 
instantaneous (pseudo) occurrence probabilities $\tilde{n}_{\pm 1/2}, n_d$, and $n_e$
for the ${\rm Mg}^+$, the ${\rm Mg}^0$, and the ${\rm Mg}^{2+}$ configuration, respectively.
Negative and positive times denote the in- and outcoming 
branch of the trajectory. The velocity $v=0.024$ and the surface temperature $T_s=400~{\rm K}$. 
Initially, the projectile is in the ${\rm Mg}^+$ configuration, that is, the lower level 
$\varepsilon_0$, representing single occupancy, is occupied while the upper level $\varepsilon_U$, 
representing double occupancy, and thus the ${\rm Mg}^0$ configuration, is empty. While the 
projectile is on its way through the trajectory the ionization levels shift and broaden. As a 
result the occupancies change. The neutralization probability is then the probability 
for double occupancy at the end of the trajectory.

For the particular case of the Mg:Au system the first ionization level, $\varepsilon_U$, that is, 
the level which has to accept an electron in order to neutralize the ion, is below
the Fermi energy of the metal throughout the whole trajectory. The broadening is also rather weak. 
It only leaks for a very short time span above the Fermi energy. As a result, the magnesium ion 
can efficiently soak in a second electron while the electron already present due to the initial 
condition is basically frozen in the second ionization level. The electron captured from the 
metal has moreover a strong tendency to stay on the projectile. It only has a chance to leave it 
in the short time span where the instantaneous broadening $\Gamma_U(t)$ is larger than 
$|{\rm E_F}-\varepsilon_U(t)|$. The neutralization probability is thus expected to be close to unity. 
Indeed, we find for the situation shown in Fig.~\ref{OccurrenceMg} $\alpha_w=n_d(\infty)=0.965$ 
(solid bullet in Fig.~\ref{OccurrenceMg}).

\begin{figure}[t]
\includegraphics[width=0.9\linewidth]{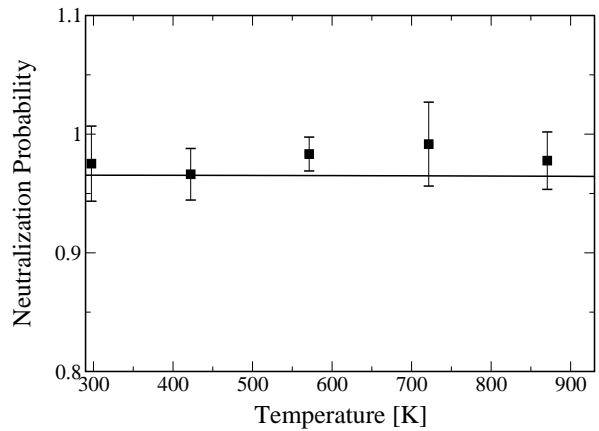}
\caption{Temperature dependence of the neutralization probability
$\alpha_w=n_d(\infty)$ for a ${\rm Mg}^+$ ion hitting with $v=0.024$
a gold surface. The turning point $z_0=5$.
Also shown are experimental data from Ref.~\onlinecite{HY11}.
}
\label{AlphaMg}
\end{figure}

The temperature dependence of $\alpha_w$ is shown in Fig.~\ref{AlphaMg}. In accordance 
with experiment we find $\alpha_w$ essentially to be independent of temperature. This 
is expected because both ionization levels, $\varepsilon_U$ and $\varepsilon_0$, are below 
the Fermi energy and their broadening is too small to allow a charge-transfer from 
the projectile to empty conduction band states of the surface. Notice, the 
excellent agreement between theory and experiment indicating that the semi-empirical 
model we developed in Sect.~\ref{Model} captures the essential features of the charge-transfer
pretty well. 

\begin{figure}[t]
\includegraphics[width=0.9\linewidth]{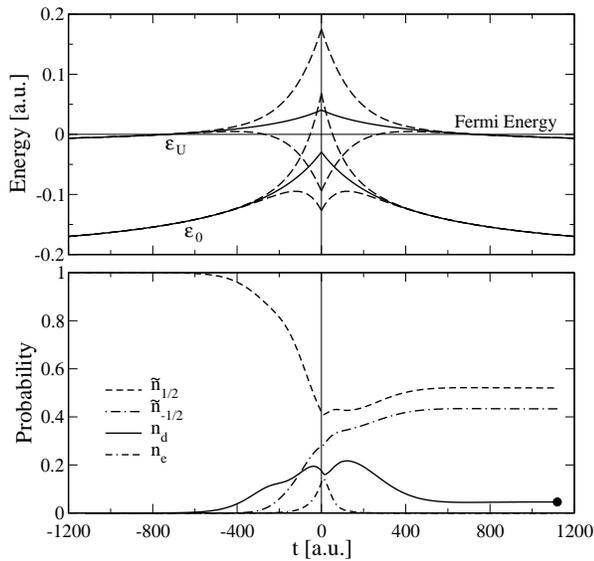}
\caption{Upper panel: Energy level diagram for the Sr:Au system at $T_s=400$~K
as a function of time. The projectile starts at $t=-1120$ and $z=z_{\rm max}=20$
with velocity $v=0.0134$, reaches at $t=0$ the turning point
$z=z_0=5$, and approaches at $t=1120$ again $z=z_{\rm max}$. The levels (solid lines)
are broadened according to $\epsilon_{0,U}\pm \Gamma_{0,U}$ (dashed lines) with
$\Gamma_{0,U}$ as shown in Fig.~\ref{LevelWidths}.
Lower Panel: Instantaneous (pseudo) occurrence probabilities along the
trajectory for the ${\rm Sr}^+$, the ${\rm Sr^0}$, and the ${\rm Sr}^{2+}$
configurations. Initially, at time $t=-1120$, the projectile is in
the ${\rm Sr}^+$ configuration. The neutralization probability
in this particular case is $\alpha_w=n_d(1120)=0.046$ (solid bullet).}
\label{OccurrenceSr}
\end{figure}

After the successful description of the Mg:Au system let us now turn to the Sr:Au system. 
In Fig.~\ref{OccurrenceSr} we again plot as a function of time the broadened ionization levels 
and the (pseudo) occurrence probabilities for the three configurations of the projectile. As 
it was the case for Mg:Au, the configuration of the projectile, which initially was in the 
configuration representing single occupancy, changes along the trajectory. The changes are 
however more subtle.

The reason is the level structure. In contrast to the Mg:Au system, the ionization 
levels are now closer to the Fermi energy of the surface. The first ionization level
$\varepsilon_U$ even crosses the Fermi energy with far reaching consequences. The part 
of the trajectory where $\varepsilon_U$ is below the Fermi energy, that is, the region 
where the neutral atom would be energetically favored, the broadening 
is very small, indicating negligible charge-transfer from the metal to the ion and 
hence a stabilization of the ion due to lack of coupling. When the broadening and thus 
the coupling is large $\varepsilon_U$ is above the Fermi level. In this part of the 
trajectory the ion is energetically stabilized. The first ionization level of 
strontium can capture an electron from the metal only in the time span where 
$|{\rm E_F}-\varepsilon_U(t)|<\Gamma_U(t)$. The neutralization probability of a 
strontium ion should be thus  much smaller than the one for a magnesium ion. Indeed 
we find $\alpha_w=n_d(\infty)=0.046$ which is much smaller than unity (solid bullet 
in Fig.~\ref{OccurrenceSr}).

Due to the shift and broadening of the first ionization level $\varepsilon_U$ it is clear 
that a strontium ion cannot as efficiently neutralize on a gold surface as a magnesium ion.
This sets the scale of $\alpha_w$. In addition, and in great contrast to magnesium, 
the second ionization level $\varepsilon_0$ is however also close to the Fermi energy. 
In those parts of the trajectory for which $|{\rm E_F}-\varepsilon_0(t)|<\Gamma_0(t)$  
it can affect the charge-transfer between the metal and the projectile. In fact, taken
by itself, it should stabilize the ion and hence decrease the neutralization 
probability.~\cite{MM98} Qualitatively, this can be understood from a density of states 
argument. From the upper panel of Fig.~\ref{OccurrenceSr} we can infer that the broadened second 
ionization level is cut by the Fermi energy in the upper half of its local density of states. 
Hence, close to the surface holes start to occupy the second ionization level at energies 
where the local density of states is higher than at the energies where electrons are 
transferred. Increasing temperature enhances thus the tendency of electron loss from the 
second ionization level. Without interference from the first ionization level the 
neutralization probability should thus go down with temperature.
\begin{figure}[t]
\includegraphics[width=0.9\linewidth]{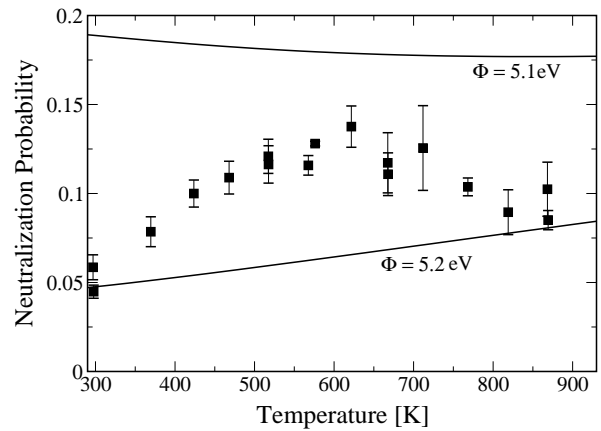}
\caption{Temperature dependence of the neutralization probability
$\alpha_w=n_d(\infty)$ for a ${\rm Sr}^+$ ion hitting with $v=0.0134$
a gold surface. The turning point $z_0=5$.
Also shown are experimental data from Ref.~\onlinecite{HY11}.
}
\label{AlphaSr}
\end{figure}

That the second ionization level of Sr comes close to the Fermi energy of Au most 
probably led He and Yarmoff~\cite{HY10,HY11} to suggest that the neutralization of strontium 
ions on gold surfaces is dominated by electron correlations. Indeed the experimentally 
found negative temperature dependence of $\alpha_w$ above $T_s=600~{K}$ seems 
to support their conclusion. However, the temperature dependence of $\alpha_w$ 
we obtain and which we plot in Fig.~\ref{AlphaSr}, does not show this behavior, 
at least, for the material parameters of Table~\ref{Parameters} and the experimentally 
measured post-collision velocity. The reason for the discrepancy between the measured
and the calculated data is unclear. The material parameters seem to be reasonable 
since the theoretical results have the correct order of magnitude. It could be
however that the temperature-induced transfer of holes to the second ionization
level is overcompensated by the electron-transfer to the first ionization level. 
In the region where charge-transfer is strongest the two ionization levels overlap.
The absence of energy separation together with the conditional temporal weighting 
due to the dynamics of the collision process makes it very hard to tell a priori 
which process will win and manifest itself in the measured neutralization probability. 

So far, the discussion of the data left out the possibility of a correlation-induced 
sharp resonance in the vicinity of the Fermi energy, that is, the key feature of
Kondo-type physics. The numerical results seem to suggest that either there is no resonance
or it does not affect the neutralization process. However, from the data itself we cannot 
determine which one is the case. We can thus not decide whether the Sr:Au system is in a 
correlated regime or not and hence whether an interpretation of the experimental data 
in terms of a mixed-valence scenario is in principle plausible or has to be 
dismissed. A rigorous way to decide this would be to calculate the instantaneous 
spectral functions for the projectile and to look for sharp resonances in the vicinity of 
the Fermi energy. This is beyond the scope of the present work.

\begin{figure}[t]
\includegraphics[width=0.9\linewidth]{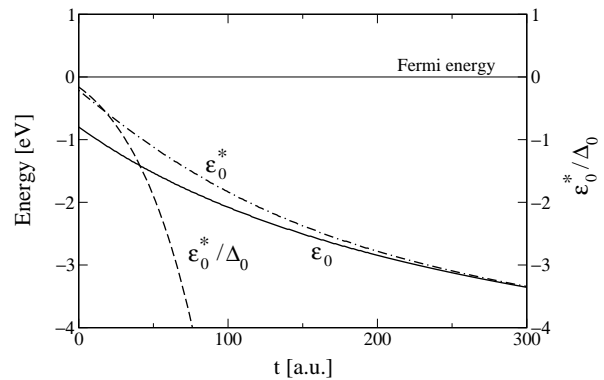}
\caption{Renormalized ($\varepsilon_0^*$) and bare ($\varepsilon_0$) second ionization
level measured from the Fermi energy as a function of time traveled along the
outgoing branch of the trajectory for the ${\rm Sr}$:Au system with $v=0.0134$
and $T_s=400~{\rm K}$. Also shown as a function of time is the scaling invariant
$\varepsilon_0^*/\Delta_0$. In the region for which $|\varepsilon_0^*/\Delta_0|<1$
the system is likely to be in the mixed-valence region. The material parameters
are as given in Table~\ref{Parameters} and $v=0.0134$.
}
\label{EstarSr}
\end{figure}

To get at least a qualitative idea about in what regime the strontium projectile 
might be along its trajectory we plot in Fig.~\ref{EstarSr}, following Merino and 
Marston,~\cite{MM98} Haldane's scaling invariant,~\cite{Haldane78} 
\begin{align}
\frac{\varepsilon_0^*}{\Delta_0}=\varepsilon_0+\frac{\Delta_0}{\pi}\log\bigg(\frac{U}{\Delta_0}\bigg)~,
\end{align}
as a function of time travelled along the outgoing branch of the trajectory. In the 
perturbative regime which is strictly applicable only far away from the surface 
$\varepsilon_0^*$ can be interpreted as the renormalized second ionization level and 
$\Delta_0=\Gamma_0/2$.~\cite{SLN94b} For $|\varepsilon_0^*/\Delta_0|<1$ the projectile 
is likely to be in the mixed-valence regime.~\cite{MM98} Since $\varepsilon_0^*$ comes 
very close to the Fermi energy holes are expected to transfer in the mixed-valence regime 
very efficiently to the projectile. In situations where the projectile stays sufficiently 
long in the mixed-valence regime before $\varepsilon_0^*$ crosses the Fermi energy 
double occupancy and hence the neutralization probability should be suppressed 
with increasing temperature. 

As can be seen in Fig.~\ref{EstarSr} close enough to the surface the strontium projectile
is indeed in the mixed-valence regime. For the material parameters given in Table~\ref{Parameters}
and the experimental value for the projectile velocity the time-span however is rather short. Most
probably this is the reason why we do not see any reduction of $\alpha_w$ with temperature for 
the parameters we think to be best suited for the Sr:Au system. Since the experimental data are 
unambiguous, this indicates perhaps the need for a precise first-principle calculation of the model 
parameters. Alternative interpretations of the experimental results can however not be ruled out.

\section{Conclusions}
\label{Conclusions}

Motivated by claims that the neutralization of strontium ions on gold surfaces is affected by
electron correlations we set up a semi-empirical model for charge-transferring collisions 
between alkaline-earth projectiles and noble metal surfaces. The surface is simply modelled 
by a step potential while the projectile is modelled by its two highest ionization levels
which couple to the surface via Gadzuk's image-potential-based projectile-surface interaction.
To calculate the neutralization probability we employed a pseudo-particle representation of
the projectile's charge states and quantum-kinetic equations for the retarded and less-than
Green functions of the projectile as initially suggested by Nordlander, Shao and Langreth. 
Besides the non-crossing approximation for the self-energies and an approximate momentum 
summation no further approximations are made. The quantum-kinetic equations are numerically 
solved on a two-dimensional time-grid using essentially the same strategy as Shao and coworkers.

The absolute values for the neutralization probability we obtain are in good agreement with
experimental data, especially for the Mg:Au system, but also for the Sr:Au system, although 
for the latter we could not reproduce the temperature dependence of the neutralization 
probability. Our calculations can thus not decide whether the He-Yarmoff experiment can be 
interpreted in terms of a mixed-valence scenario. From the instantaneous values of 
Haldane's scaling invariant we see however that the Sr:Au system could be in the mixed-valence 
regime. The mechanism for a negative temperature dependence, that is, the possibility of 
efficiently transferring holes to the second ionization level, is thus in principle present. 
For the material parameters however most appropriate for Sr:Au the negative temperature 
dependence arising from this channel seems to be overcompensated by the positive temperature
dependence of the electron-transfer to the first ionization level. To proof that He and 
Yarmoff have indeed seen--for the first time--mixed-valence correlations affecting charge-transfer 
between an ion and a surface requires therefore further theoretical work.

\section*{Acknowledgements}
M. P. was funded by the federal state of Mecklenburg-Western Pomerania through a 
postgraduate scholarship within the International Helmholtz Graduate School for Plasma 
Physics. In addition, support from the Deutsche Forschungsgemeinschaft through project 
B10 of the Transregional Collaborative Research Center SFB/TRR24 is greatly acknowledged.


\end{document}